\documentclass[12pt]{article}

 \usepackage{amsfonts}   \usepackage{amssymb,amsmath}
 \usepackage{graphicx} \usepackage{epstopdf}
\newcommand{\be}{\begin{equation}}
\newcommand{\ee}{\end{equation}}
\newcommand{\ba}{\begin{eqnarray}}
\newcommand{\ea}{\end{eqnarray}}
\newcommand{\bi}{\begin{itemize}}
\newcommand{\ei}{\end{itemize}}
\newcommand{\la}{\label}
\newcommand{\non}{\nonumber}
 
 \newcommand{\crlb}[1]{\label{#1}\\[2pt]}
 
 \newcommand{\crld}[1]{\label{#1}}
 \newcommand{\eela}[1]{\quad\hbox{\scriptsize{#1}}\label{#1}\end{eqnarray}}
 \newcommand{\eelb}[1]{\label{#1}\end{eqnarray}}
 
 \newcommand{\newsecb}[2]{\section{#1}\label{#2}\setcounter{equation}{0}}

 \newcommand{\nolabels} {\def\eel{\eelb} \def\crl{\crlb} \def\newsecl{\newsecb}\def\bibiteml{\bibitem}\def\citel{\cite}\def\labell{\crld}}

\newcommand\publishversion{\nolabels\setlength{\textheight}{8.75in}\setlength{\oddsidemargin}{0in}
    \setlength{\textwidth}{6.3in}\setlength{\topmargin}{-0.3in}}

        \def\be{\begin{eqnarray}}    \def\ee{\end{eqnarray}}
 \def\bi#1{\begin{itemize}\item[#1]}     \def\ei{\end{itemize}}  
   \def\^#1{\hat{#1}}

 \def\d{\delta}      \def\D{\Delta}   
       \def\l{\lambda}      \def\m{\mu}
                 \def\n{\nu}
             \def\r{\varrho}     \def\s{\sigma}  
 \def\t{\tau}          
               \def\ch{\chi}
 \def\w{\omega}

 \def\ffract#1#2{\raise .2 em\hbox{$\scriptstyle#1$}\kern-.3em/
                 \kern-.2em\lower .15 em \hbox{$\scriptstyle#2$}}

\def\bmatrix{\begin{matrix}} \def\ematrix{\end{matrix}} \def\bpmatrix{\begin{pmatrix}}\def\epmatrix{\end{pmatrix}}
\def\bcenter{\begin{center}} \def\ecenter{\end{center}}

\def\lowerheightfig#1#2#3{\(\raise-#1\hbox{\includegraphics[height=#2]{#3}}\)}
\def\lowerwidthfig#1#2#3{\(\raise-#1\hbox{\includegraphics[width=#2]{#3}}\)}

\usepackage{color}

 \publishversion

\begin{document}

\bcenter 
{\LARGE\textbf{Raymond and instantons: some recollections
and the use  of ADHM\\[25pt]}}
{\large Chris P. Korthals Altes}  \\[20pt]
National Institute for subatomic physics NIKHEF\\[5pt] Theory group\\[5pt]

Science Park 105\\[5pt] 
1098 XG, Amsterdam\\ [5pt]

The Netherlands  \\[10pt] 

and\\[10pt]

Centre de Physique ThŽorique\\[5pt]
Campus de Luminy, Case 907\\[5pt]
163 Avenue de Luminy\\[5pt]
13288 Marseille cedex 9, France\\[10pt]

\vfil
{\bf In memory of Raymond Stora}\\
1930 -- 2015
\ecenter
\vfil
\begin{quotation} \noindent {\large\bf Abstract } \medskip \\
After the discovery of the BRST identities in 1974 Raymond spent some two years mostly on instantons.  In those years
we had a small group at the Centre Physique Th\'eorique in Marseille discussing the physics and mathematics of instantons. 
 The upshot of our discussions can be found in a set of lectures given by Raymond in Erice in 1977 and a year later in a Physics Reports volume.  I  present some recollections of that period; mostly how 
 we were influenced by the twistor approach.  I discuss the Atiyah-Drinfeld-Hitchin-Manin (ADHM) method to obtain instantons  ("calorons") in thermal QCD following earlier work.  The building blocks are a prepotential and a gauge invariant propagator.
 Then I will give  surprisingly simple results in terms of these building blocks for the one loop vacuum response to a change in caloron parameters and some physical consequences for the effective action, in particular screening properties.
 
 \end{quotation}
\vfil 
\noindent Version: April 2016 \   \\[2pt]
{\footnotesize Last typeset: \today}
\eject
\setcounter{page}{2} 
\section{Introduction}\la{sec:intro}

On a spring morning in 1976 Raymond walked into my office at the CPT waving a paper written by a quartet of Russians.  It was the celebrated solution to the Yang-Mills equations in vacuo, the  "pseudo-particle" solution, now known as instanton,
 by Belavin, Polyakov, Schwartz and Tyupkin~\cite{polyakov}.  
Within a week or so he had formed a group of interested people, including John Madore, Jean Louis Richard and myself.   Raymond was fascinated not so much by 
the physics of instantons but more by the geometry behind them. Geometry means here : algebraic geometry.  In that sense a boutade by Bethe applies to him. Bethe said famously of Oppenheimer that "Oppenheimer did physics because he thought it was the best way of doing philosophy".  Raymond  loved quantum field theory because he may have thought this was the best way to do mathematics.  If true, QFT did profit in profound ways from his attitude!

The reason Raymond had been alerted on the Russian paper was probably the discovery early in 1976 by 't Hooft~\cite{gth1} that the pseudo-particles in the presence of quarks 
became sources for the anomalous chiral current $j^5_\m$ 
\be
\partial_\m j^5_\m={g^2\over{4\pi^2}}tr F\tilde F.
\ee
Here $\tilde F$ is the dual of the field tensor~\footnote{$ \tilde F_{\m\n}={1\over 2}\epsilon_{\m\n\r\s}F_{\r\s}$, $E_k=F_{4k}$, $B_k=\epsilon_{klm}F_{lm}$}.
In fact the space time  integral of the divergence is the change in $Q^5=\int d\vec x j^5_0(x)$ and can be an even integer through the Pontryagin number Eq. (\ref{topcharge}).
That had profound implications for the axial $U(1)$ breaking, the structure of the groundstate~~\cite{gross}\cite{jackiw} and hence phenomenology. The configurations with non-vanishing Pontryagin number are not small fluctuations from the trivial, or pure gauge configuration, as
in perturbation theory. Hence it was a departure from perturbation theory and the beginning of a new era in QCD where non-perturbative 
physics became accessible in a quantitative way.

However we left this developments aside. We were  interested in the construction of  multi-instantons and their degeneracy,  and embarked on the twistor program by Penrose. The twistor program was seen by Atiyah's group in Oxford (consisting of H.J. Hitchin, R.S. Ward, joined by I.M. Singer and many others) as the way to make inroads into the mathematics of the instanton. 
There was a twistor letter edited by this group every few months showing a feverish activity.
I dug out such a letter from my notes from that year 1976, it contained handwritten notes by Penrose.

In Marseille we had regular meetings, sometimes twice a day.  The main thrust was to establish the number of  gauge inequivalent instanton solutions (the moduli space) as a function of its  topological charge or 
Pontryagin number
 
\be
Q=\int_{R_4} d^4x tr F\tilde F=k {8\pi^2\over{g^2}}, ~k~ \mbox{integer}.
\la{topcharge}
\ee
Any self- or antiself dual configuration ($\vec E=\pm \vec B$) is a solution of the equations of motion. This follows from the Bianchi identity
 
 \be
 D_\m\tilde F_{\m\n}=0.\non
 \ee
 
 Hence such a configuration is a local extremum, but in fact the following identity shows it is a local minimum: 
  \be
{1\over 2} \int dt d\vec x (F\pm\tilde F)^2=S\pm Q\non.
 \ee
 
 As the left hand side  is a square the action $S=\int dt d\vec x F^2$ is never smaller then the absolute value of the topological charge $Q$. Only if the field is (anti)-self dual there is equality, and there is a minimum.  The equations are first order differential equations, still a formidable challenge.

The twistor approach had to do with complex null planes in four dimensional complex Euclidean space. Consider two vectors in such a plane with the property that their norms and inner product were zero
\be
\xi^2=\eta^2=\xi.\eta=0.\non
\ee 

 At the same time Penrose supposed the van der Waerden representation $\overline {\xi}$ of  such vectors can be written as the direct product 
of two complex two-spinors
\ba
\overline{ \xi}\equiv \xi_0+i\vec\t.\vec\xi&=&\pi\otimes v \t_2\non\\
\underline{{\xi}}\equiv\xi_0-i\vec\t.\vec\xi&=&v\otimes\pi\t_2\non.
\la{penrose}
\ea
The spinor $\pi$ characterizes the plane and is defined up to a complex multiplicative number. The other spinor
is specific to the vector.  This Penrose representation 
selects a handedness  $\chi$ for the ($\xi,\eta$) plane because $\underline{\xi}\overline{\eta}$ contains a factor $\pi\t_2\pi$, so vanishes
\be
0=\underline{\xi}\overline{\eta}=i\vec\t(-\vec \xi\eta_4+\vec\eta\xi_4+\chi \vec\xi\wedge\vec\eta)\non
\la{handedness}
\ee
\noindent with $\chi=1$.
And so the field tensor with $\vec E=\vec B$ vanishes in this plane
\be
F_{\xi\eta}=F_{\m\n}\xi_\m\eta_\n=\vec E.(-\vec \xi\eta_4+\vec\eta\xi_4)+\vec B. \vec \xi\wedge\eta=0.
\la{twistor}
\ee

After an early paper by Ward~\cite{twistorward}  Atiyah and Ward showed~\cite{twistorwardatiyah} in 1977
that minimum action solutions for SU(2) Yang-Mills fields in Euclidean 4-space correspond, via this Penrose twistor transform, to algebraic bundles on the complex projective 3-space. These bundles in turn correspond to algebraic curves. 

 The projective space was wetting Raymond's  appetite in the subject.


In the mean time  methods based on supersymmetric 
 arguments~\cite{brown1} produced the dimension of the moduli space rather easily: 8k. And 't Hooft (unpublished, 1976) found a quite simple form for an instanton potential with charge $k$, size $\r_l$ and location at $x_{(l)}$\footnote{From now on I write $({\bf 1},i\vec \t)~\mbox{as}~ \s_\m=\overline{\s}_\m, ~\mbox{and}~\s^\dagger_\m =\underline{\s}_\m, \epsilon_{1234}=1,~\mbox{and} ~\eta_{\m\n}=\s_{[\m}\s^\dagger_{\n]}={1\over 2}(\s_\m\s^\dagger_\n-\s_\n
 \s^\dagger_\m)$}.
 He noticed that a gauge potential of the form
 \be
 A_\m&=&{1\over 2}\bar \eta_{\m\n}\partial _\n\log  \phi(x)
 \ee
 was self dual if $\phi(x)$ is quasi-harmonic
\be
 {1\over \phi(x)}\Box\phi(x)=0.\non
 \ee
 Then, if $\phi(x)$ is finite at infinity then it is fixed (up to an overall constant)
 \be
  \phi(x)&=&1+\sum_{l=1}^k {\r_l^2\over{(x-x_{(l)})^2}}. \la{gthk}
 \ee
 
 This solution has $4k$ translation and $k$ scale degrees of freedom, stlll not the total of $8k$ solutions.
 It has singularities at the locations of the instantons.  But they can be removed by a gauge transformation.

 Two years later Atiyah, Drinfeld, Hitchin and Manin published their definitive  ("ADHM") paper~\cite{ADHM} on the construction of instantons.
 It generalizes 't Hooft's solution in a simple way.
 It was certainly definitive in a mathematical sense: they showed that there is a very simple Ansatz for {\it any} instanton and that this Ansatz avoids solving first order partial differential equations. One needs to solve an algebraic equation. This was a big step forward. But solving the 
 algebraic equations in specific cases asked for very insightful guesswork by the (mathematical) physicist.  That insight was provided in specific cases by beautiful work of Nahm~\cite{nahm}, Lee~\cite{lee} and van Baal~\cite{pierre2} some years later.  In section \ref{sec:caloron} the reader will get an idea of what was involved. 
 { Those familiar with the classical aspects of calorons may skip Section \ref{sec:caladhm} and start with the sections on quantum corrections.}

  Meanwhile in Marseille I had turned to more
 mundane aspects of instantons. The others  continued the approach with twistors, Raymond produced the Erice lectures~\cite{storaerice} in 1977, and the three of them published a beautiful review on 
 twistor methods in 1978~\cite{twistorstora}. 
 
  Ironically it was only 35 years later that I returned to  instantons, as I had got involved in a Festschrift on instantons in the high temperature version of QCD.  I wanted to understand quantitatively what  instantons (aptly called calorons in the high temperature context)  contributed to the free energy.  The method of ADHM is then an unavoidable tool to get to the classical caloron solution. And to my surprise the very quantities
  that are the cornerstones of their approach survive in a simple way in the quantum fluctuations around the classical solution.  In what follows 
  I would like to tell this story.   
  I feel somewhat guilty when using this beautiful mathematics in the mundane context of QCD. After all it was Hardy who categorized maths into "beautiful but useless" and "useful but ugly".  I hope, perhaps naively, that I have combined the beautiful and the useful. 
 
 \section{Caloron and ADHM}\la{sec:caloron}\la{sec:caladhm}

In this section I give a short description of the ADHM method for the construction of an instanton with topological charge  $k$. The gauge group will be $SU(2)$.
{I will use the same conventions as Kraan and van Baal in their seminal paper~\cite{pierre2}.}
It starts by writing the gauge potential  in terms of a prepotential $N$
   \be
A_\m=N^\dagger\partial_\m N.
\la{prepot}
\ee

The prepotential  is a $k+1$ column vector with real quaternionic entries and is normalized
\be
N^\dagger N=1.
\la{normN}
\ee

 Then clearly the potential is anti-Hermitean.
 
 $SU(2)$ gauge transformations $g$ act identically on every entry in the prepotential, $N\rightarrow Ng$. That gives
the familiar transformation law for the vector potential in (\ref{prepot}), using (\ref{normN}).

The upper entry is chosen to be $-1$. This fixes the gauge.  The $k$ lower 
entries are denoted by $u_l, l=1,\cdots,k$:
\be
N={1\over \phi^{1/2}}{\left( \begin{array}{c}
 -1\\
 u_1\\
 \vdots\\
 u_k
  \end{array} \right)} 
 \la{Ncolumn}.
  \ee
Here $\phi$, a c-number, is chosen to normalize $N$, see (\ref{normN}).

 If the instanton has charge $k$ $M$ is a matrix with k+1 rows and k columns. 
Like for the prepotential every entry is a real quaternion, i.e.
\be
 M_{ij}^\m\s_\m
 \la{Mcoeff}
 \ee
 \noindent with real coefficients. 
The first row has  k entries $\l_1, \l_2, \cdots \l_k$. They are quaternionic generalizations of the k size parameters in 't Hooft's Ansatz (\ref{gthk}).  

The fundamental Ansatz is that the remaining $k\times k$ matrix contains all the $x$ dependence~\footnote{From now on the quaternions $x=x_\m\s_\m, x_n=n{\bf 1}-x$ are used and $x_n^2=x_n x _n^\dagger$ is a real number.} and only linearly 
\be
B- x{\bf 1}_{k\times k}
\ee
\noindent or
\be
M={\left( \begin{array}{ccc}
 \l_1&\cdots&\l_k\\
 B_{11}-x&\cdots& B_{1k}\\
  \vdots&\ddots&\vdots\\
 B_{k1}&\cdots& B_{kk}-x
  \end{array} \right)}. 
 \la{Mmatrix}
  \ee


The prepotential is in the orthogonal complement of  the matrix $M$
\be
M^\dagger N=N^\dagger M=0.
\la{complement}
\ee

%
  The ADHM construction is completed by the key requirement that $M^\dagger M$ is proportional to ${\bf 1}$.   
     So the product is free of the three Pauli matrices, hence implies  three quadratic constraints in terms of the
  real coefficients constituting a quaternionic matrix element of $M$, as in (\ref{Mcoeff}).  
  
   That the field tensor is then  indeed self-dual follows from a nice property of the covariant derivative $D=\partial+N^\dagger\partial N$ using
   the normalization of $N$ and the Leibnitz rule
   \ba
   D_\m N^\dagger&=&\partial_\m N^\dagger+N^\dagger\partial_\m N N^\dagger\non\\
   &=&\partial_\m N^\dagger\left({\bf 1}-N N^\dagger\right).\non
   \ea 
   Now ${\bf 1}-N N^\dagger$ annihilates $N$ and is a projector due to the normalization (\ref{normN}). The matrix $M({M^\dagger M})^{-1}M^\dagger$ has the same properties due to (\ref{complement}), so the two are the same. In what follows we write for the inverse
   \be
   R=(M^\dagger M)^{-1}.\non
   \la{inverse }
   \ee

After substitution and using  (\ref{complement}) once more, the Leibnitz rule and the linearity of $M$ in $x$ we get for the covariant derivative
 \ba
 D_\m N^\dagger&=&- N^\dagger\partial_\m M RM^\dagger\non\\
 &=&N^\dagger \s_\m R M^\dagger.
 \la{covariant2}
 \ea
This form of the covariant derivative of $N$ serves as corner stone for the classical {\it and} quantum instanton calculations.
 
 The field tensor becomes after substitution of (\ref{covariant2}) 
    \ba
  F_{\m\n}&=&D_\m N^\dagger\partial_\n N-{\m\leftrightarrow \n}\non\\
  &=&N^\dagger \s_\m R M^\dagger\partial_\n N-{\m\leftrightarrow \n}.\non
\ea
Once more using $M^\dagger\partial_\n N=-\partial_\n M^\dagger N$ and the fundamental Ansatz (\ref{Mmatrix}) for $M$
\ba
F_{\m\n}
&=&N^\dagger\s_\m  R\s^\dagger_\n N-{\m\leftrightarrow \n}.\non
\ea
As $R$ is a $k\times k$ matrix it acts only on the lower components of $N$ or $N^\dagger$ in Eq. (\ref{Ncolumn}). So from (\ref{Ncolumn}) 
follows $RN=\phi^{-1/2}Ru$\footnote{The self duality of the field tensor is a general result, valid without the restriction (\ref{Ncolumn})}. 

Since the propagator $R$ is gauge invariant it commutes with the $\s$'s,
and the field strength becomes proportional to 't Hooft's self dual tensor 
\ba 
F_{\m\n}
&=&{1\over{\phi}}u^{\dagger}(\s_\m \s^\dagger_\n-\s_\n\s^\dagger_\m) R u\non\\
&=&{2\over{\phi}}u^{\dagger}\eta_{\m\n}R u.
\la{fieldstrengthdual2}
\ea

So this proves that any gauge field configuration obeying the Ansatz  is indeed self-dual. That this Ansatz gives {\bf all} $SU(2)$ instantons with charge $k$ is the amazing result of the ADHM paper.

 By straightforward algebra one obtains once more for the gauge potential~\cite{pierre2} 
\ba
A_\m&=&{1\over{2\phi}}(u^\dagger\overleftrightarrow{\partial}_\m u)\non\\
&=&{\phi\over 2} \l \bar\eta_{\m\n}\partial_\n R\l^\dagger. 
\la{gaugepotR}
\ea

$R$ can now  simply  be related to the propagator $S=(B^\dagger-x^\dagger)(B-x))^{-1}$ by noting that (\ref{prepotsource})  implies through the normalization $u^\dagger u=\phi-1$
\be
u^\dagger u=\l S\l^\dagger=\phi -1.\non
\la{Sphi}
\ee
It follows that~\footnote{$\l (\l^\dagger)$ is a row (column) vector as defined in (\ref{Mmatrix}). So $(S\l^\dagger\l S)_{mn}=S_{mk}\l^\dagger_k\l_l S_{ln}$}
\be
R=S-{1\over \phi}S\l^\dagger \l S.
\la{RS}
\ee
The latter equality acting on $\l^\dagger$ provides us with a simple  relation between the two propagators:
\be
R\l^\dagger={1\over \phi} S\l^\dagger.\non
\ee
This relation leads with (\ref{Sphi}) to another useful relation
\be
\l R\l^\dagger=1-{1\over \phi}
\la{Rphi}
\ee
Using  (\ref{Sphi}) once more one finds the second expression for the vector potential in (\ref{gaugepotR}).

The action density was computed ~\cite{osborn} in these terms and turns out to be
\be
tr F^2_{\m\n}=-\partial^2_\m\partial^2_\n\log\det R.
\la{actiondensity}
\ee
The integral over space-time is then given as in Eq. (\ref{topcharge}). It only depends on the topological charge, not on  size or other parameters\footnote{This will change
in the quantum corrections!}.

Because of (\ref{gaugepotR}) and (\ref{actiondensity}) the propagator $R$ is indeed the  centerpiece of the construction.

The 't Hooft Ansatz (\ref{gthk}) is recovered by a {\it diagonal} B matrix with  the $l$th diagonal element the position
{ $x_l\equiv (x_{(l)\m}-x_\m)\s_\m$} in quaternion form, and as scale factor $\l_l=\r_l g$, $\r_l$ a positive real number. 

Let's recapitulate. One first has to guess the form of the source $\l$ and of  the matrix $B$. To have a gauge invariant propagator one has to solve quadratic equations in terms  of the entries in $M$. Then (\ref{complement}) can be written as 
\be
(B^\dagger-x^\dagger)^{-1}\l^\dagger=u.
\la{prepotsource}
\ee
  
From the prepotential $u$ follows the gauge potential through (\ref{prepot}) and (\ref{Ncolumn}). Alternatively the gauge potential can be expressed as a matrix element of the propagator $R$ in terms of the sources $\l$. For the field strength one needs both the prepotential $u$, and   the propagator $R$.    
\subsection{Calorons, classical}
In hot QCD  one can find  thermal  instantons by using the periodicity modulo $1/T$ of Euclidean space. The simplest is the periodic instanton\cite{hs}  obtained from the 't Hooft
Ansatz  (\ref{gthk}), by taking the single instanton, located say at $x_{(n)}=0$, and repeated in the time direction. What follows assumes $T=1$. The source  $\l_n=\r$, the matrix $B$ is diagonal,  $B_{nn}-x=n-x\equiv x_n$, and $x_nx_n^\dagger\equiv x^2_n$
\ba
A_\m&=&{1\over 2}\bar\eta_{\m\n}\partial_\n\log \phi,\non\\
\phi(x)&=&1+\sum_n u_n^\dagger(x)u_n(x)\non\\
u_n(x)&=&(x_n^\dagger)^{-1}\r={x_n\over{x_n^2}}\r.
\la{periodicinstanton}
\ea 
The propagator $R$ follows simply   from the known form of $B$ and using  (\ref{RS})
\be
R_{m, n }={1\over {x_m^2}}\d_{mn}-{1\over \phi}{\r^2\over{x_m^2x_n^2}}.
\ee
The index $n$ runs now from $-\infty$ to $\infty$. The topological charge is unity, when we limit ourselves to a single time slice. Note that every individual tem in $\phi(x)$
drops off quadratically in the space direction. But the thermal  sum over all slices gives a $\phi(x)$ that drops off only linearly. This will give rise to thermal screening, see the discussion in section \ref{sec:freeenergy}.

However the story does not end here! 

The free energy in hot QCD (strictly speaking gluodynamics)  depends on an order parameter,  the trace of the path ordered Polyakov loop 
\be
P=tr {\cal P}\exp(\int_0^{1/T}A_4(\vec x,t)dt).
\ee

Under periodic gauge transformations this order parameter is invariant.

In our case we suppose the loop asymptotes at spatial infinity to
\be
\exp(i2\pi \w \t_3).
\la{asyloop}
\ee
The isotopic 3-direction  is a matter of choice.
\begin{figure}   
  \centering 
  \includegraphics{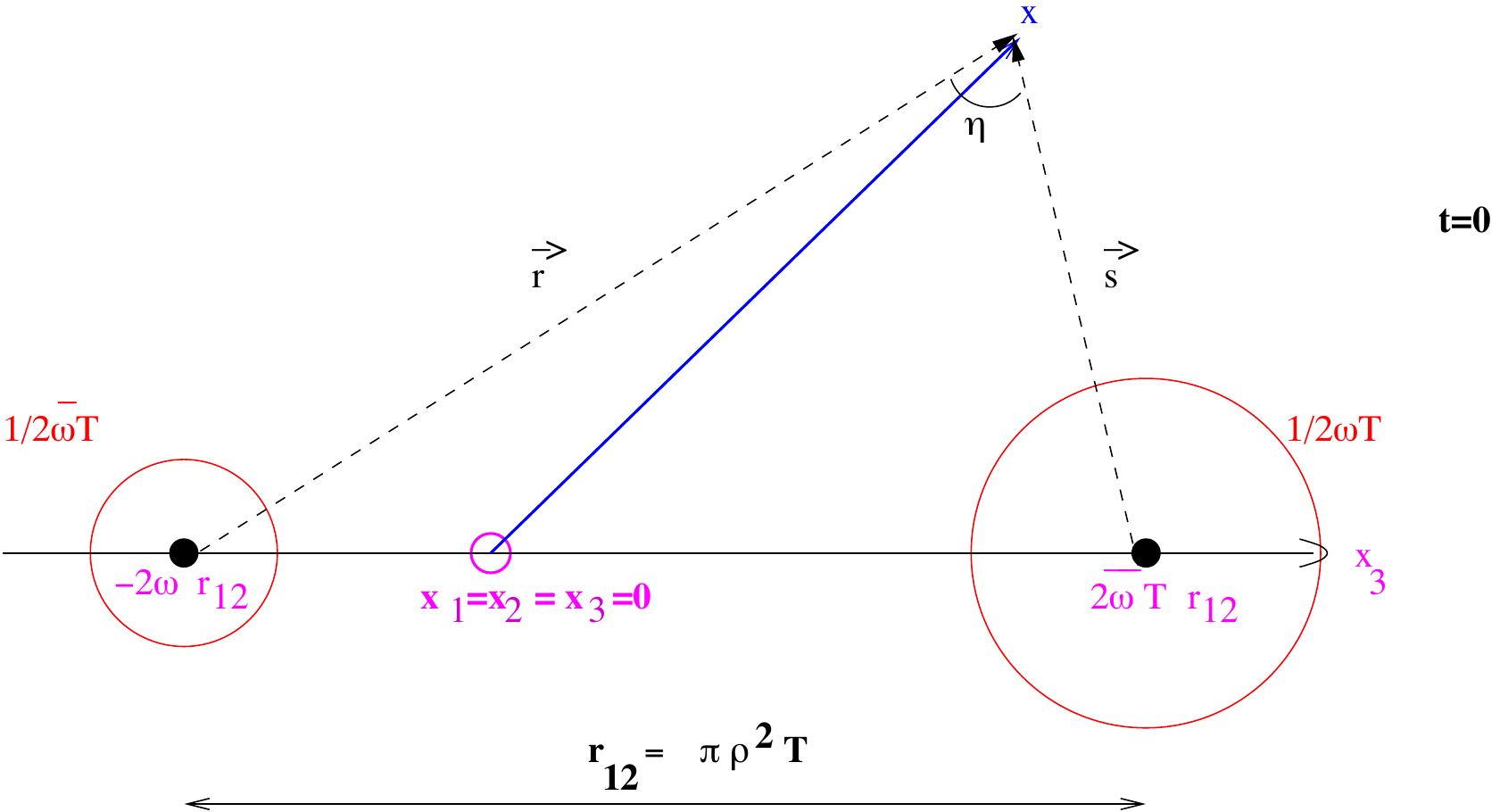}
  \caption{The caloron with two monopoles, with centers and cores as in the text. Their separation $r_{12}$ ( Eq. (\ref{separationinstantsize})) is taken to be larger than the cores.}
  \label{fig:caloron}
 \end{figure}     

 At very high temperature
the one loop  free energy $f$is periodic mod 1/2 and  has two degenerate minima, one at $\w=0$, and one at $\w=1/2$,
\be
f(T, \w)=\pi^2 T^4\left(-{1\over 15}+{16\over 3}\w^2(1-2|\w|)^2\right)
\la{freeen}
\ee
The $\w$ dependence is given by that in the Bernoulli polynomial 
\be
B_4(x)=\sum_{n\neq 0} -{\cos(2\pi nx)\over{(2\pi n)^4}}
\la{bernoulli1}
\ee

 In both minima we have a gas of free  gluons, and the value of the loop
is $\pm 1$.  The symmetry of the system is the center  symmetry  $Z_2$  of the gauge group $SU(2)$, and this symmetry is spontaneously broken at high temperature. When the symmetry is restored the order parameter will vanish, which happens at $\w=1/4$, which stays the location of the minimum at and below the transition temperature. This symmetry maps $\w$ into $1/2-\w\equiv\bar\w$~\footnote{The potential has a non-convex part which in this mean field approach leads to negative squared masses, see section \ref{sec:quantumcaloron}}.

The question is now whether instantons with the order pararmeter different from $\pm 1$ ("calorons") exist and how they influence the free energy in between very large $T$ and the critical temperature.  To find the solution it is quite helpful to realize that the instanton in this very narrow Euclidean time slice (i.e. $1/T<\r$, $\r$ the size of the single instanton in a time slice) is almost a particle, and that it becomes
a bound state~\cite{lee}  of two self-dual BPS 't Hooft-Polyakov monopoles, whose Higgs field is $A_4$. The latter asymptotes into the VEV
\be
A_4=i {\t_3\over 2}\left(4 \pi \w+\left({1\over r}-{1\over s}\right)\right) +\mbox{exponentially damped core terms}.
 \la{asy}
\ee 
This is consistent with (\ref{asyloop}) 
{and describes the long range dipole field  of the caloron.}
The first Coulomb term 
   is   the due to the static BPS monopole, with topological charge and mass $2\w$ and a core of size $(2\w)^{-1}$. The second  one is due to a BPS monopole transformed by   a  gauge transform periodic modulo $Z(2)$, with opposite magnetic charge,  topological charge and mass $2\bar\w$ and core size ${(2\bar\w)}^{-1}$. The bound state they form has topological charge $1$ and no net magnetic or electric charge.  The dimensionality of the individual monopole moduli space
 is four, three from the translational modes plus  one internal mode. So the dimension of the moduli space for the composite matches that of the instanton with charge $\pm 1$: dimension $8$.
 
 This should be  the solution of the  caloron for large separation of the constituent monopoles.  To find its explicit solution for all separations we fix the four  translational modes at the origin, and the three internal rotational modes at unity.  What is left is the size of the caloron which is taken to be the separation $r_{12}$~\cite{lee}\cite{pierre1}\cite{pierre2}. So  viewed from a field point $\vec x$
 one monopole is at position 
 \be
  \vec r=\vec x +2\w r_{12}\vec e_3, \mbox{the other at}~ \vec s=\vec x-2\bar\w r_{12}\vec e_3
  \la{rscms}
  \ee
  \noindent  in the center of mass system as in Fig. (\ref{fig:caloron}).  This is true for all times $x_4$ for very large separation $r_{12}$. Sure, for small separation and fixed $T$ we will find back the single  instanton localized in time.

   We will now  find the ADHM source $\l$ and propagator $R$ by sticking to this parametrization,  even for small separation.

The first, natural, guess was that the instanton size $\r$ carries the holonomy $\exp(i2\pi \w\t_3)$, instead of being the same in each time slice.
So in the $n$th time slice
\be
\l_n=\r\exp(i 2\pi n \w \t_3).
\la{sizecal}
\ee
But how to find the matrix $B$?  

This task is much simplified by changing from the labeling by the integers $n$ to the  Fourier transform to the circle
\be
\l(z)=\sum_n\l_n\exp(i2\pi n z)=\r\left({1+\t_3\over 2}\d(\w+z)+{1-\t_3\over 2}\d(\w-z)\right).
\la{sourcez}
\ee

So the source becomes on the circle  a sum of two Dirac delta functions at $z=\pm \w$. In the inverse propagator  $M^\dagger M$
its square appears again as a sum of Dirac delta functions but with the square of the strength $\r^2(1\pm\t_3)/2$.

 The $B$ matrix on the circle takes the simplest form possible consistent with $M^\dagger M$ being gauge invariant.
 
  It has a kinetic term in $z$ due to the diagonal element $(n-x)\d_{mn}$, familiar from the periodic instanton (\ref{periodicinstanton}). The off-diagonal  elements  must furnish   a potential which has to be piece wise constant in order to match the delta functions from $\l(z)$ in $M^\dagger M$ we discussed before 
\be
(B-x)(z,z')= \left({\partial_z\over{2\pi i}}-x_4 -\vec\s.\vec s \ch_{[-\w,\w]}(z)-\vec\s.\vec r\chi_{[\w,1-\w]}(z)
\right)\d(z-z').
\la{Bcaloron}
\ee

With this choice the inverse propagator $M^\dagger M$ becomes
\ba
M^\dagger M(z,z')&=&\left((B-x)^\dagger(B-x))+\l^\dagger\l\right)(z,z')\non\\&=&\left(\left({\partial_z\over{2\pi i}}-x_4\right)^2+r^2\chi_{[\w,-\w]}(z)+s^2\chi_{[-\w,\w]}+{\r^2\over 2}\left(\d(z-\w)+\d(z+\w)\right)
\right)\d(z-z')\non\\ &+&\left({r_{12}\over{2\pi}}-{\r^2\over 2}\right)(\d(z+\w)-\d(z-\w))\t_3\d(z-z')
\la{caloronpropagator}
\ea
The kinetic term in $B^\dagger$ produces delta functions from the piece wise constant potential of $B$, and they are shown in the last line, together with the $\t_3$ dependent terms in $(\l^\dagger\l)(z)$.
The propagator has to be proportional to the unit quaternion.  Hence the monopole separation is fixed in terms of the single instanton size $\r$ by
\be
r_{12}=\pi\r^2.
\la{separationinstantsize}
\ee

The separation is in terms of the {\it square} of the single instanton size  because of the quadratic constraint.

  The propagator $R$ becomes on the circle
the inverse of the Schroedinger equation (\ref{caloronpropagator}), which describes a particle on the circle in a potential
with two repulsive delta functions of equal strength and in between a potential mountain of height $\vec r^2-\vec s^2$.

So (\ref{Bcaloron}) and (\ref{caloronpropagator}) give us the gauge potential and self-dual field strength through (\ref{fieldstrengthdual2}) and
(\ref{prepotsource}).

\subsection{Long range behavior}\la{longrange}

The caloron has indeed the long range behavior of a pair of monopoles, if their separation exceeds the cores.  The long range fields of the
monopoles are in the $\t_3$ direction, due to our choice of the asymptotics in (\ref{asy}).

To see this we split the second expression for the vector potential in (\ref{gaugepotR}) into a part where $\bar\eta$ commutes with the source $\l$
and a remainder. The source is proportional to the projectors $P_\pm\sim 1+\t_3$ so commutes with $\bar\eta^3\t_3$. The remainder will be orthogonal 
to $\t_3$. So the long range behavior we are after is entirely contained in the first part
\ba
A_\m&=&i{\phi\over 2}\bar\eta^3_{\m\n}\t_3 \l\partial_\n R\l^\dagger\non\\
&=&i\bar\eta^3_{\m\n}{\t_3\over 2}\partial_\n\log\phi+~\mbox{core suppressed}.
\la{a4}
\ea
The last equality follows from (\ref{Rphi})~\footnote{ $\bar\eta^a_{\m\n}$ is the 't Hooft symbol with $\bar\eta^3_{43}=1$}.  
From the definition of the field strengths we find easily
\be
B^3_k=E_k-\triangle\log\phi, ~E_k=\partial_k\partial_3\log\phi
\ee 

What is  the long range behavior of $\phi$? That turns out to be quite simple if $\w\bar\w\neq 0$, because then we can drop all reference to the cores and 
\be
\phi={r+s+r_{12}\over{r+s-r_{12}}}+~\mbox{core suppressed}.
\ee 
In this approximation $\log\phi$ is harmonic, $\triangle \log\phi=0$, except on the segment in between the monopoles where the denominator vanishes. There a Dirac string develops. But the Dirac string reduces to two three dimensional delta functions in $\triangle A_4$. From (\ref{a4})  and the properties of $\bar\eta$ we get
\be
\triangle A_4^3=\partial_{x_3}\triangle\phi=-4\pi\left(\d(x_3+2\bar\w r_{12})-\d(x_3-2\w r_{12})\right)\d(x_1)\d(x_2)
\ee 
\noindent hence the Coulomb potentials in $A_4$, Eq. (\ref{asy}), follow.

But the {\it full} solution (\ref{gaugepotR}) has only one singular point at the origin, as behooves an instanton. The monopoles
are regular at the center of their respective cores, and this avoids the Dirac string.

This ends our discussion of the classical caloron, except for a practical remark.

The solution is not periodic as the choice of the source $\l(z)$ already indicated.  However, in  thermal field theory one needs a periodic background. To render the solution periodic we apply  a non-periodic gauge transformation 
\be
g=\exp(i2\pi x_4\w\t3)\non
\ee
 on $N$. After transformation the prepotential gets as upper component in (\ref{Ncolumn})
\be
-\exp(i2\pi x_4\w \t_3) 
\la{wilsonlineprepot}
\ee
whilst the lower components become
\be
u(x,z)g=w(x,z),
\la{periodicprepot}
\ee 
 \noindent with $w(x+l,z)=w(x,z)\exp(i2\pi l z), l~ \mbox{integer}$. In this gauge the scalar potential $A_4$ approaches  $2\pi i\w \t_3$ at spatial infinity as in Eq.(\ref{asy}) .
 
 If the instanton size $\r$ vanishes this scalar potential is the only term that survives. So the caloron can be viewed as  a self-dual dipole superposed on this constant background. 
 
 In what follows this  caloron potential in periodic gauge  is used. Vanishing holonomy $\w=0$ reduces it to the periodic instanton (\ref{periodicinstanton}).

\section{Quantum effects of the caloron, and their simplicity}\la{sec:quantumcaloron}

The quantum effects of  any instanton in the semi-classical approximation can be expressed in terms of the fluctuation determinant of a scalar isovector particle~\cite{gthdet}. In terms of the gauge covariant Klein Gordon operator $D(A)^2$:
\be
\log \det -D(A)^2
\ee
't Hooft~\cite{gthdet} computed this determinant for the single instanton.  
 In the same year as the ADHM paper a series of important papers by Lowell Brown and coworkers appeared\cite{brown1}\cite{brown2}. Its subject was the propagators of massless particles in the background of an instanton. Its aim was to calculate the quantum effects of the many-instanton\cite{brown3}, generalizing  the single instanton calculation by 't Hooft~\cite{gthdet}. Other papers on the same subject but using ADHM~\cite{christ}\cite{corrigan}\cite{osborn}\cite{nahm} appeared soon after. In the context of thermal QCD  periodic instanton determinants were computed by Gross et al.~\cite{pisarski}. Zarembo~\cite{zarembo} computed the fluctuation determinant  for a single constituent of the caloron, while Diakonov et al.~\cite{diakonov} and Korthals Altes et al.~\cite{sastre} analyzed the caloron.
 
We will concentrate in this section on the polarization current ${\cal J}_\m$ one obtains by varying one of the parameters of the caloron gauge potential $A_\m$. With the variation $\d A_\m$ the response of the system is driven by the polarization current ${\cal J}_\m$
\ba
\d\log \det -D(A)^2&=&\int^{1/T}_0 dt  d\vec x Tr_c\d A_\m {\cal J}_\m(x)\non\\
{\cal J}_\m(x)&=&\overrightarrow{D}_\m \Delta_p(x,y)+\Delta_p(y,x)\overleftarrow{D}_\m
\la{polarizationcurrent}
\ea

As we said above, periodicity in the gauge potential is assumed and
\ba
\overrightarrow{D}_\m&=&\partial_\m+A_\m\non\\
\overleftarrow{D}_\m&=&-\partial_\m+A_\m.
\ea

The Hermitean conjugate of the propagator is supposed to be same but with its arguments exchanged. If so the polarization current is anti-Hermitean, like the vector potential.

 The periodized propagators are constructed from the traditional ones by 
\be
\D_p(x,y)=\sum_n\D(x,y+n), ~\mbox{with}~ -D^2\D(x,y)=\d(x-y).
\ee
The propagator for an isospin 1/2 particle is quite simple~\cite{christ} and consists of the overlap of the prepotentials $N(x)$
\be
\Delta(x,y)={N^\dagger(x)N(y)\over{4\pi^2(x-y)^2}}.
\la{iso1/2prop}
 \ee
 Due to the normalization of the prepotentials, Eq (\ref{normN}),  this expression reduces to the free propagator for coinciding points.
To obtain the polarization current 
we  take the limit $x\rightarrow y$ in Eq. (\ref{polarizationcurrent}).  As long as $n\neq 0$ the terms are finite and adding all those  we get  the "thermal" polarization current\footnote{For the $n=0$ component see references~\cite{corrigan}.}. So we split the current into an $n=0$ part ${\cal J}^{(0)}_\m$ and the rest ${\cal T}_\m$, the thermal part on which we will concentrate. It will turn out to be a quite simple expression in terms of the prepotentials $w$ and the propagator $R$.

The idea will be to use again the covariant derivative of $N$, as we did in the previous section, Eq.(\ref{covariant2}), to obtain the field tensor.  To evaluate the polarization current, Eq. (\ref{polarizationcurrent}), all we need is
\ba
D_\m N^\dagger(x)N(y)
 &=&N^\dagger(x) \s_\m R(x) M^\dagger(x) N(y)\non\\
 &=&N^\dagger(x) \s_\m R(x) \left(M^\dagger(x)-M^\dagger(y)\right) N(y)\non\\
&=&n N^\dagger(x) \s_\m R(x) N(x+n).
 \la{covariantNN}
 \ea
To get the last line the linearity in x and y from the fundamental Ansatz (\ref{Mmatrix}) is used and of course the periodicity $(y-x)_\m=n\d_{4,\m}$.  

The thermal part of the polarization current takes then a  simple form, using (\ref{covariantNN}) and the relation between left and right derivatives in (\ref{polarizationcurrent}). It consists of  {two} parts, one where the covariant derivatives act on the prepotentials and one part where the derivative acts on the 
free propagator $1/(x-y)^2)$. The latter contributes only to the charge density since $(y-x)_\m=\d_{\m,4}n$.

\ba
{\cal T}_\m(x)&=&\sum_{n\neq 0}\Bigg[{1\over{4\pi^2 n}} \left(N^\dagger(x) \s_\m R(x) N(x+n)-N^\dagger(x+n) R(x)\s^\dagger_\m N(x)\right)\non\\&+ &{\d_{\m 4}\over {2\pi n^3}}( N^\dagger(x)N(x+n)-N^\dagger(x+n)N(x))\Bigg].
\la{polarizationcurrent2}
\ea
As before $R N= \phi^{-1/2}R w$ since $R$ acts only on the lower components of $N$. Not surprisingly, given the way we get the vector current (the first term in Eq. (\ref{polarizationcurrent2})), this term is a point split version of the  field tensor $F_{\mu 0}$ (see (\ref{fieldstrengthdual2})). 

As we work in periodic gauge we should take Eq. (\ref{wilsonlineprepot}) and (\ref{periodicprepot})   into account and after some simple rearranagement the result is
\ba
{\cal T}_\m(x)&=&\sum_{n\neq 0}\Bigg[{1\over{4\pi^2 n}}\int_0^1dz dz' \left(w^\dagger(x,z) \s_\m R(z,x,z') w(x,z')\exp(i2\pi n z')-h.c.\right)\non\\&+ &i{\d_{\m 4}\over {\pi n^3}}\left( \sin(2\pi\w n)\t_3+\int_0^1 dz w^\dagger(x,z)w(x,z)\sin(2\pi \w n z)\right)\Bigg].
\la{polarizationcurrent3}
\ea

\begin{figure}   
  \centering 
  \includegraphics{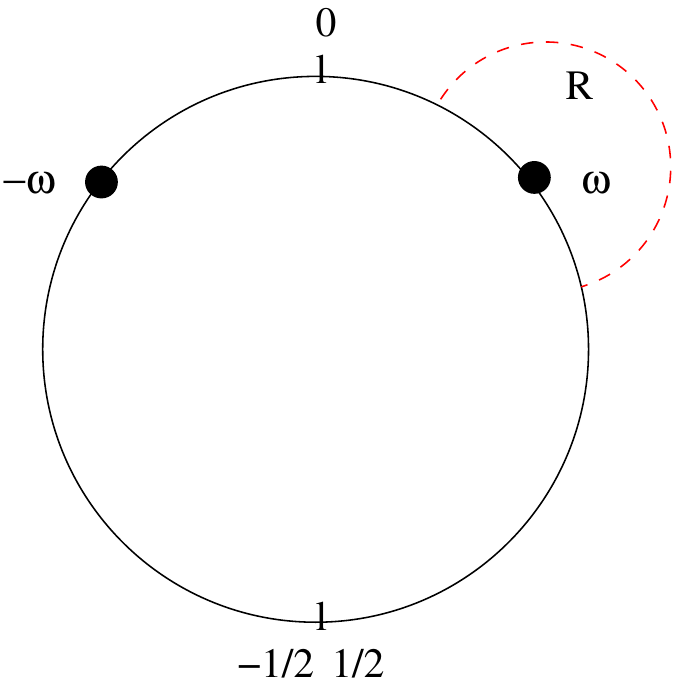}
  \caption{The R propagator connects on the circle the segment $[-\w,\w]$ and  its complement.}
  \label{fig:Rprop}
 \end{figure}

So the vectorial part  of the current is expressed in double integrals over the circle of the propagator $R$, $w$ and $w^\dagger$. This is reflecting the fact that the R propagator   connects the prepotentials $w(x,z)$ on  the two segments, see  {Fig.(\ref{fig:Rprop})}. The charge density gets a constant contribution $\sin(2\pi n\w)\t_3$ which originates in the non-trivial holonomy
and equals the imaginary part of the Wilson line product $\exp(-i2\pi \w x_4)\exp(i2\pi\w(x_4+n)\t_3)$. The time dependence has therefore dropped out. 

We can do the sum over $n$ trivially as
\ba 
\sum_{n\neq 0}{\sin(2\pi n z )\over{(2\pi n)^3}}&\equiv &B_3(z)={1\over 12} (2 z^3-3 z^2\epsilon(z)+ z)\\
\sum_{n\neq 0}{\sin(2\pi n z )\over{2\pi n}}&\equiv &B_1(z)= (-z+ {1\over2}\epsilon(z)),
\la{sumovernz}
\ea
\noindent the well known Bernoulli polynomials for odd integer $k$,  periodic modulo one, with the  property $B'_{k}(z)=(-)^{k-1}B_{k-1}(z)$. 

Using the expression for $w(x,z)$  from  Eq. (\ref{prepotsource})  and  (\ref{periodicprepot}) one easily integrates the overlap $w^\dagger(x,z)w(x,z) B_3(z)$ over the circle. The result for the interval $[-\w,\w]$ 
is that the Bernoulli polynomial $B_3(z)$ is reproduced  as function of $\w$ and three other terms proportional to the derivatives of this Bernoulli polynomial, accompanied by higher 
powers in $1/s$:
\be
\int_{-\w} ^\w dz w^\dagger(x,z)w(x,z)B_3(z)=\sum^3_{k=1}{1\over s^k}B_{4-k}(\w)W(\hat s, s).
\la{powerrsorderB}
\ee
The quaternionic matrix  $W(\hat s, s)$ contains apart from the directional vector $\hat s= {\vec s\over s}.\vec \t$ only s dependence through the core.  

 Similar for the interval $[-\bar\w, \bar\w]$. So the long distance behaviour is related to $\w$ derivatives of the free energy of the bulk, Eq (\ref{freeen}), using (\ref{bernoulli1}).  Note the appearance of the non-convexity of the potential Eq. (\ref{freeen}) for $k=2$.   

For the vectorial current the  propagator $R(z,x,z')$ requires additional effort.

%
%

\subsection{The determinant and the effective action}\la{sec:freeenergy}

Let us now consider the variation of the determinant and consider only the thermal part
\be
{\d A_\m.{\cal T}_\m}=\d A_4.{\cal T}_4+\d A_l .{\cal T}_l.
\la{detcal}
\ee
{The dot indicates integration over the thermal space-time slice and color tracing.} 

In its generality the resulting integrations over the circle and then over space time do not deliver something immediately interesting. But it is interesting
to look at the long distance properties, i.e.
at what happens when neglecting the cores in the variation of $A_\m$. Then, as we saw,  only the 4- component with its Coulomb terms and the azimuthal component 
survive. We look at the variation in $\w$ and obtain for the first  term in (\ref{detcal}) dropping core terms in the potential
\ba
\d A_4.{\cal T}_4&=&\int^{1/T}_0 dx_4\int d\vec x \left(2\pi\d\w+\d_{\w}\left(1/r-1/s\right)+...\right) \non\\&\times&\left(B_3(\w)+tr_c\t_3\int^1_0 w^\dagger(x,z)\w(x,z)B_3(z)+..\right).
\la{coulomb}
\ea

Remember that the prepotential is proportional to $\r$,
{~from Eq. (\ref{prepotsource}) and (\ref{sizecal})}.  The thermal current is bilinear in the prepotentials apart from the holonomy term $B_3(\w)$.
So if we let the distance $r_{12}=\pi \r^2$ vanish only the holonomy term will survive. By the same token only the VEV will survive in the vector potential.
So in that limit all what remains is the variation of the VEV and the holonomy term  and their product
gives us a volume term. 
This volume term is the variation of the free energy of the constant Wilson line background, which was calculated long ago~\cite{pisarski}.   
 
Now we address the question: how does the Coulomb force change through the fluctuations? One would anticipate Debye screening through the terms in the overlap we discussed
below  Eq. (\ref{sumovernz}). Traditionally Debye screening is expected when the electric fields of the caloron are stretched through the separation $r_{12}$ becoming large. This is what happens for the caloron with trivial holonomy~\cite{pisarski} and the resulting screening is {\it linear} in $r_{12}$. This was one of the motivations
to look at instantons at high temperatures, since the screening renders the integration over their size infrared finite. 

However something amusing and unexpected happens due to the presence of the non-trivial holonomy term  $B_3$ in  ${\cal T}_4$. 
The variation of the Coulomb force in (\ref{coulomb}) combined with this holonomy term  gives a screening term proportional to  the  {\it square} of the monopole distance $r_{12}$! This is due to obvious dimensional reasons when doing the integral over space.
The variation itself is proportional to $r_{12}$ according to Eq. (\ref{rscms}).  And the integral of the variation of $1/r$ equals apart from this factor $r_{12}$ ($r^2_{\bot}=x_1^2+x_2^2$)
\be
\int d\vec x {x_3+2\w r_{12}\over{((x_3+2\w r_{12})^2+r_{\bot}^2)^{3/2}}}.
\ee  
  It produces after integration over $r_{\bot}$ a sign function in $x_3$ which is centered at the corresponding monopole position.
The variation of $1/s$ gives a sign function centered in the monopole position with opposite charge.  The difference between the two cancels the wings of both
sign functions, and the resulting $x_3$ integral equals $2r_{12}$, the distance between the monopoles. The result for the  determinant after integration over the holonomy becomes
\be
8\pi r_{12}^2(B_4(\w)-B_4(0)).
\la{isoonehalfresult}
\ee

So we wind up with a quadratic screening term\footnote{ There is an apparent ambiguity: doing the $x_3$ integration first would give a vanishing result, since the integrand is odd in $x_3+2\w r_{12}$ and shifting the integration variable seems permitted. However it is not permitted in each of  the sign functions apart, as they are linearly divergent.}. From earlier work\cite{pisarski, sastre} we know it should vanish at zero holonomy.

Up till now we discussed the isospin $1/2$ determinant. The isospin  one determinant is more involved, but the quadratic screening term is of the same form as in Eq. (\ref{isoonehalfresult}), but with the argument $\w$ of the Bernoulli function replaced by $2\w$.

Note  that the quadratic screening extends into the region where the potential is non-convex. So the quadratic screening  might mean that we are entering an unstable branch of the free energy. 

We have not yet finished the calculation of the other subleading screening effects but expect them to be of the form  in ref.~\cite{diakonov}, due to Eq. (\ref{powerrsorderB}).   

\section{Epilogue}\la{sec:epi}

These notes are meant to be an illustration of how a mathematical construction like ADHM originally describing minima of the classical Yang Mills action can be of direct use for the corresponding one loop correction to the effective action. The result is a polarization current that is determined in terms of a propagator and a prepotential obtained from simple Schroedinger potentials on the circle. The appearance of a quadratic screening term is amusing and deserves further scrutiny.  

In March 2015  we saw  the Stora's  at their  home in Saint Jean de Gonville.  I told him I had been revisiting the instanton mathematics of the seventies. He was amused and recalled how he had heard about the ADHM construction from Atiyah,  back in 1978. 

Raymond remains for us the master of  welding together  the serenity of  mathematics and the beauty of physics.

\end{document}